\documentclass[aps,prb,twocolumn,showpacs,amsmath,amssymb]{revtex4}
\usepackage{epsfig}
\usepackage{bm}
\def\be{\begin{equation}}
\def\ee{\end{equation}}

\begin{document}
\title{Dynamic generation of orbital quasiparticle entanglement in mesoscopic conductors} 
\author{P. Samuelsson}\thanks{Present adress: Departement of Solid State Theory, Lund University, S\"olvegatan 14 A, S-223 62 Lund, Sweden}
\author{M. B\"uttiker} 
\affiliation{D\'epartement de
Physique Th\'eorique, Universit\'e de Gen\`eve, CH-1211 Gen\`eve 4,
Switzerland} \date{\today}

\begin{abstract}
We propose a scheme for dynamically creating orbitally entangled
electron-hole pairs through a time-dependent variation of the
electrical potential in a mesoscopic conductor. The time-dependent
potential generates a superposition of electron-hole pairs in two
different orbital regions of the conductor, a Mach-Zehnder
interferometer in the quantum Hall regime. The orbital entanglement is
detected via violation of a Bell inequality, formulated in terms of
zero-frequency current noise. Adiabatic cycling of the potential, both
in the weak and strong amplitude limit, is considered.
\end{abstract}

\maketitle 

Entanglement is an aspect of quantum mechanics which has intrigued and
fascinated physicists for seven decades.\cite{Schrod} While the
initial interest mostly concerned the interpretation and fundamental
properties of quantum mechanics,\cite{EPR,Bell} recently entanglement
has emerged as a resource for quantum information and computation
purposes. A controlled creation, manipulation and detection of
quasiparticle entanglement in mesoscopic conductors is therefore of
interest. In particular, computation is a time-dependent process with
elementary steps often driven by a masterclock. \cite{rlmb} Clearly it
would be desirable to be able to produce entanglement on command, once
per clock-cycle.

Here we propose, as a first step towards time-controlled quasiparticle
entanglement, a scheme for the dynamic generation of orbitally
entangled \cite{sam1} quasiparticle pairs in a mesoscopic
conductor. Orbital entanglement\cite{sam1,sam2,bss,qh1,qh2} uses
quasiparticles generated by two spatially separated sources. The
spatial index of the sources plays the role of a pseudo-spin
index. Due to indistinguishability of the pair creation events, the
two sources generate entangled electron-electron\cite{sam1} or
electron-hole \cite{qh1} states. The main advantages of
orbital entanglement are that local rotations of the orbital single
particle state can be performed with experimentally
available\cite{schon,oliver} electronic beamsplitters and that detection, in
contrast to spin entanglement, can be done without spin-to-charge
conversion.

\begin{figure}[b]   
\centerline{\psfig{figure=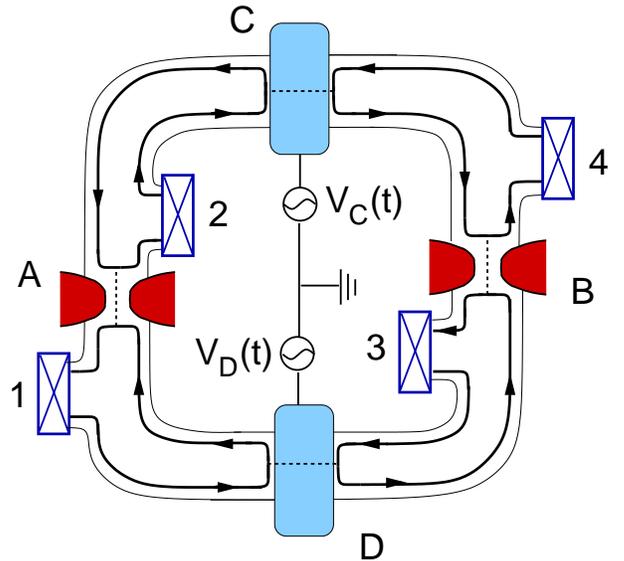,width=8.0cm}}   
\caption{Mach-Zehnder geometry in the quantum Hall
  regime. Transport takes place along a single edge state (thick black
  line) in the direction shown by the arrows. The potentials $V_C(t)$
  and $V_D(t)$ at $C$ and $D$ are adiabatically and periodically
  modulated in time, creating electron-hole pairs propagating towards
  reservoirs 1 to 4. The static point contacts at $A$ and $B$ work as
  controllable beam-splitters.}
\label{fig1}
\end{figure}

In the set-up considered here the sources are provided by electrical
potentials that are varied periodically and adiabatically in two
spatially separated regions in the conductor, a Mach-Zehnder
interferometer in the quantum Hall regime (see Fig.~\ref{fig1}).  The
oscillating potentials excite electron-hole pairs, orbitally entangled
with respect to the regions of emission. The emitted quasiparticles
are detected in electronic reservoirs, all kept at zero bias. Our
scheme bears some resemblance to quantum pumping
\cite{pumping,Mosk,Mosknoise,pumpnoise} but due to the chiral nature
of the geometry no net electrical current per clock-cycle is
generated.  Instead the excited quasiparticles give rise to current
noise. Shot noise in the absence of dc-current has recently been
measured in an experiment which excites quasiparticles with the help
of an oscillating contact potential.\cite{glattli} We show that the
entanglement can be detected via a Bell Inequality formulated in terms
of the period-averaged, zero frequency noise. Although the scheme is
non-ideal, i.e. it generates less than one entangled pair per cycle,
it is a simple and transparent proposal of entanglement generated by
time-dependent electrical potentials. Moreover, the geometry is
similar to a Mach-Zehnder interferometer recently realized
experimentally,\cite{ji} making our proposal experimentally highly
relevant.

There are to date a large number of suggestions for creation of
spin\cite{dot,scond} as well as orbital\cite{sam1} entanglement in
quantum dot,\cite{dot} superconductor,\cite{scond,sam1,scond2} quantum
Hall\cite{qh1,qh2,sam2} and beam-splitter \cite{bs,leb1} systems. A common
feature of the majority of the proposed schemes is that the entangled
quasiparticles are emitted in a random and uncontrolled fashion, by
e.g. tunneling across a potential barrier. This is similar to the
creation of entangled photon-pairs by parametric
down-conversion\cite{Zeil} in optics. Although this form of
entanglement is useful for basic quantum information processes, for
more complex tasks such as quantum computation it is however desirable
to have a time-controlled source of entangled quasiparticles.

In our proposal for dynamic generation of entanglement (see
Fig. \ref{fig1}) we consider a Mach-Zehnder interferometer in the
quantum Hall regime, connected to four reservoirs, $1$ to $4$, in
thermal equilibrium.  Transport takes place along a single,
spin-polarized edge state. Due to the chiral nature of the edge
state-transport,\cite{halp} backscattering\cite{mb88} takes place only
at the scatterers $A,B,C$ and $D$. At $A$ and $B$ quantum point
contacts work as controllable beam-splitters.\cite{schon} At $C$ and
$D$ (electrostatic top- or side gates) the corresponding electrical
potentials $V_C(t)$ and $V_D(t)$ are modulated periodically in time,
with a period $\tau=2\pi/\omega$. The adiabatic limit is considered,
where the amplitudes for scattering between the reservoirs $1$ to $4$
are independent on energy on the scale of the clock frequency
$\omega$. Only results to first order in $\omega$ are discussed and
the temperature is taken much smaller than $\hbar\omega$.

We point out that quasiparticle entanglement in static, voltage-biased
quantum Hall systems have been investigated previously in
Refs. [\onlinecite{qh1,qh2,sam2,bss}]. Here we instead consider a situation
where all reservoirs are kept at zero bias but the scattering
potentials for the electrons in regions $C$ and $D$ are
time-dependent. To consider the simplest possible system,
quasiparticles are, in contrast to Refs. [\onlinecite{qh1,qh2,sam2,bss}],
injected and detected at the same reservoirs. This however makes the
scheme in Fig. \ref{fig1} less nonlocal.

Due to the oscillating potentials at $C$ and $D$, electrons incident
from the reservoirs 1 to 4 can absorb or emit one or several quanta
of energy $\hbar \omega$ before propagating out to the reservoirs
again. In this Floquet picture\cite{Mosk,Mosknoise} the scattering
in both energy and real space can be described by scattering matrices
\begin{equation}
S_{C/D}(E_n,E)=\left(\begin{array}{cc} r_{C/D}(E_n,E) & t'_{C/D}(E_n,E) \\  t_{C/D}(E_n,E)  & r'_{C/D}(E_n,E) \end{array}\right)
\end{equation}
where e.g. $t_C(E_n,E)$ is the amplitude for an electron incoming at
energy $E$ from left towards $C$ to be transmitted to the right at
energy $E_n=E+n\hbar\omega$. The dependence of the scattering
amplitudes on $V_{C/D}(t)$ is determined by the properties of the
scattering potential in the regions $C$ and $D$, here we only work
with the scattering amplitudes themselves to keep maximum generality.

We first focus on the limit of weakly oscillating potentials,
$V_{C/D}(t)=V_{C/D}+\delta V_{C/D}\cos(\omega t+\phi_{C/D})$, with
$\delta V_{C/D}$ so small that only the amplitudes to absorb or emit
one quanta need to be taken into account. The relevant scattering
amplitudes are then e.g. $t_C\equiv t_C(E,E)$ and $\delta
t_C^{\pm}=\delta t_C\mbox{exp}(\mp i\phi_C)\equiv t_C(E_{\pm 1},E)$,
with $\delta t_C=\delta V_C(\partial t_C/\partial V_C)/2$, and similar
for the other amplitudes. Formally the weak potential limit, $|\delta
t_C|^2 \ll 1$ etc, implies that the scattering amplitudes are weakly
dependent on energy on the scale of the potential variations $\delta
V_{C/D}$.

To demonstrate that entanglement is created by the oscillating
potentials, we first consider the state of the particles emitted from
the two regions $C$ and $D$. It can be constructed from the many-body
state of the electrons incident from the reservoirs $1$ to $4$,
(suppressing spin)
$|\Psi_{in}\rangle=\prod_{j=1}^{4}\prod_{E}a_j^{\dagger}(E)|0\rangle$,
where $|0\rangle$ is the true vacuum and $a_j^{\dagger}$ creates an
electron at energy $E$, incident from reservoir $j$. Introducing
operators, e.g. $b_{AC}^{\dagger}(E)$ creating an outgoing electron at
energy $E$ at contact $C$ propagating towards contact $A$, we can
relate the $b$-operators to the $a$-operators at $C$ as
\begin{equation}
\left(\begin{array}{c} b_{AC}(E) \\
  b_{BC}(E)\end{array}\right)=\sum_{n=0,\pm 1}
  S_C(E,E_n)\left(\begin{array}{c} a_{2}(E_n) \\
  a_{4}(E_n)\end{array}\right)
\label{barel}
\end{equation}
and similarly at $D$. Inserting these relations into $|\Psi_{in}\rangle$
and expanding to first order $\delta V_{C/D}$, we find the state
outgoing from $C$ and $D$ in terms of the $b$-operators as
\begin{equation}
|\Psi_{out}\rangle=|\bar 0 \rangle+\int_{-\hbar
 \omega}^0dE\left(|\Psi_{out}^C(E)\rangle+|\Psi_{out}^D(E)\rangle\right)
\label{stateexp}
\end{equation}
with 
\begin{equation}
|\Psi_{out}^C(E)\rangle=\sum_{\alpha,\beta=A,B}f_{\alpha\beta}^C 
b_{\alpha C}^{\dagger}(E_1)
b_{\beta C}(E)|\bar
0\rangle
\label{stateexpC}
\end{equation}
where $f_{AA}^C=\delta r_C^+r_C^*+\delta t'^+_Ct_C'^{*},~f_{BB}^C =
\delta t_C^+t_C^*+\delta r'^+_Cr_C'^{*},~f^C_{AB}=\delta
r_C^+t_C^*+\delta t'^+_Cr_C'^{*},~f^C_{BA}=\delta t_C^+r_C^*+\delta
r'^+_Ct_C'^{*}=-e^{-2i\phi_C}(f^{C}_{AB})^*$ and
$|\Psi_{out}^D\rangle=|\Psi_{out}^C\rangle$ with $C\rightarrow D$. The
ground state in terms of outgoing operators is $|\bar
0\rangle=\prod_{E}b_{AC}^{\dagger}(E)
b_{BC}^{\dagger}(E)b_{AD}^{\dagger}(E)
b_{BD}^{\dagger}(E)|0\rangle$. Each term in Eq. (\ref{stateexpC})
contains an operator product $b_{\alpha C}^{\dagger}(E_1)b_{\beta
C}(E)$ acting on $|\bar 0\rangle$, describing the destruction of one
electron at an energy $-\hbar\omega<E<0$ below the Fermi surface,
i.e. the creation of a hole, and the creation of an electron at energy
$0<E_1<\hbar\omega$ above (in leads $\alpha C$ and $\beta C$
respectively).
\begin{figure}[b]   
\centerline{\psfig{figure=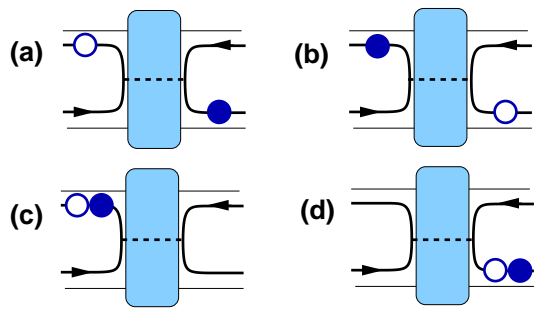,width=7.0cm}}   
\caption{The four different electron-hole pair emission processes at
  $C$: In the two upper processes, the pair is split and (a) the
  electron is emitted towards B and the hole towards A, or (b) the
  electron towards A and the hole towards B. In the two lower
  processes, both quasiparticles are emitted towards (c) A or (d)
  towards B. The same processes occur at $D$.}
\label{fig2}
\end{figure}
The effect of the weak potential oscillations is thus
to create electron-hole pair excitations out of the ground
state.\cite{Moskcom} The four different scattering processes at each
contact $C$ and $D$ are shown schematically in Fig. \ref{fig2}.

The excited state in Eq. (\ref{stateexp}) is clearly orbitally
entangled, it is a linear superposition of electron-hole pair
wavepackets emitted at $C$ and $D$, i.e. the contact indexes $C$ and
$D$ form an orbital two-level system, or qubit. In first quantization,
for identical scattering amplitudes at $C$ and $D$, the state can be
written $(|CC\rangle+|DD\rangle)\otimes |\bar \Psi\rangle$, where
$|CC\rangle+|DD\rangle$ is maximally entangled, an orbital Bell state,
and $|\bar \Psi\rangle$ contains all additional properties of the
state, e.g. energy dependence and quasi-particle character. We note
that the emitted wave-packets contain quasiparticles in the energy
range $\hbar \omega$. The clock frequency thus, in this respect, plays
the same role as the applied voltage in the entanglement schemes in
e.g. Refs [\onlinecite{sam1,sam2,qh1,qh2,bs}]. However, due to the
absence of an applied bias there is no need to reformulate the ground
state to reveal the entanglement, as was done in
Ref. [\onlinecite{sam1,sam2,qh1,qh2}].

The entanglement is detected via cross-correlations of electrical
currents flowing in leads $1,2$ (at A) and $3,4$ (at B). A 
number of works have investigated the noise properties of pumped
mesoscopic conductors, see
e.g. Refs. [\onlinecite{pumpnoise,Mosknoise}]. 
The operators $b_j$ for outgoing electrons at
reservoirs $j=1$ to $4$ are related to the $b_{\alpha\beta}$ operators
via the (energy-independent) scattering matrices of the static,
controllable point contact at $A$ as
\begin{equation}
\left(\begin{array}{c} b_{1}(E) \\
  b_{2}(E)\end{array}\right)=\left(\begin{array}{cc} \cos\theta_A &
  \sin\theta_A \\  -\sin\theta_A  & \cos\theta_A \end{array}\right) \left(\begin{array}{c} b_{AC}(E) \\
  b_{AD}(E)\end{array}\right)
\end{equation}
and similarly at $B$. The current operator is $I_j(t)=(e/h)\int
dEdE'\mbox{exp}(i[E-E']t/\hbar)[b^{\dagger}_j(E)b_j(E')-a^{\dagger}_j(E)a_j(E')]$.
The average current $\langle I_j(t) \rangle$ contains in general a dc-
and an ac-part. However, the dc-current in the geometry under
consideration is zero, i.e. there is no dc-electrical current, only
quasiparticle current is generated. This can be understood
qualitatively from the fact that the scattering amplitudes of the
injected electrons, as a consequence of the chiral transport, only
depend on one scattering potential, at $C$ or $D$. Since adiabatic
one-parameter pumps\cite{pumping,Mosk} do not pump any net current,
the dc current in our geometry is zero. The ac-current at contact $j$
is proportional to the amplitude for both quasiparticles in an
electron-hole pair emitted at $C$ or $D$ to be scattered to reservoir
$j$. This provides however no information about the entanglement.

The cross-correlation between the currents $i=1,2$ and $j=3,4$,
averaged over the time difference $t'$, is given by
\begin{equation}
S_{ij}(t)=\int dt' \langle \Delta I_i(t) \Delta I_j(t+t')+\Delta I_j(t+t')\Delta I_i(t)\rangle
\end{equation}
with $\Delta I_j(t)=I_j(t)-\langle I_j(t) \rangle$. The noise, just as
the current, has a dc and an ac-part,
i.e. $S_{ij}(t)=S_{ij}^{dc}+S_{ij}^{ac}(t)$. The dc-part is, for
e.g. reservoirs $1$ and $3$
\begin{eqnarray}
S_{13}^{dc}&=&\frac{e^2}{\tau}\left[|f_{AB}^C\sin\theta_A\sin\theta_B+f_{AB}^D\cos\theta_A\cos\theta_B|^2
  \right.
\nonumber \\
&&\left.+|f_{BA}^C\sin\theta_A\sin\theta_B+f_{BA}^D\cos\theta_A\cos\theta_B|^2\right]
\label{currcorr}
\end{eqnarray}
This expression has a simple physical explanation. The first term in
the bracket in Eq. (\ref{currcorr}) is the amplitude for an emitted
electron-hole pair from $C$ or $D$ to split, with the electron ending
up in reservoir $1$ and the hole in $3$. The second term is just the
amplitude for the opposite process, the electron detected in $3$ and
the hole in $1$. The other correlators $S_{ij}^{dc}$ are found
similarly. We note that only the scattering processes where the pair
splits [(a) and (b) in Fig. \ref{fig2}] contribute to the leading
order cross-correlators.\cite{sam1}

Considering for simplicity the case where the two scattering
potentials at $C$ and $D$ are equal, i.e. $t_C=t_D=t$ etc, up to the
pump phases $\phi_{C/D}$, one can write
\begin{eqnarray}
S_{13}^{dc}&=&S_0\left[\cos^2\theta_A\cos^2\theta_B+\sin^2\theta_A\sin^2\theta_B
  \right. \nonumber \\
  &&\left.+2\gamma\cos\theta_A\cos\theta_B\sin\theta_A\sin\theta_B\right]
\label{currcorr2} 
\end{eqnarray}
with $S_0=(2e^2/\tau)|\delta rt^*+\delta t'r'^*|^2$, $\gamma=\cos
\varphi \cos(\phi_C-\phi_D)$ and $S^{dc}_{24}=S^{dc}_{13}$ and
$S^{dc}_{14}=S^{dc}_{23}=S^{dc}_{13}(\theta_B \rightarrow
\theta_B+\pi/2)$. Here $\varphi$ is an overall phase containing
possible scattering phases of contacts $A$ and $B$ and phases due to
propagation along the edge states, including Aharonov-Bohm phases. The
noise correlator is proportional to $|\delta rt^*+\delta t'r'^*|^2$, 
i.e. proportional to $\delta V^2$. The last term in the bracket, the
interference term, is proportional to $\cos(\phi_C-\phi_D)$, i.e.it is
maximized for the two pumping potentials in phase. Due to the
phase-dependent term $\cos \varphi$, the noise correlators show a
two-particle Aharonov-Bohm effect, similarly to
Ref. [\onlinecite{sam2,bss}]. It is found that the ac-noise provides
no further information about the entanglement.

The orbital entanglement in Eq. (\ref{stateexp}), focusing on the
split quasiparticle pairs detectable via the cross correlators, is
independent on whether the electron is emitted towards $A$ and the
hole towards $B$ [process (a) in Fig. \ref{fig2}] or vice versa
[process (b)].  A Bell Inequality \cite{Bell,clau} can be formulated
in terms of the probability to jointly detect \cite{sam1,sam2} one
quasiparticle at $A$ and one at $B$ during a clock-cycle. This
probability is formally defined as
$P_{ij}=\int_0^{\tau}dtdt'P_{ij}(t,t')$ with
$P_{ij}(t,t')=P_{ij}^{eh}(t,t')+P_{ij}^{he}(t,t')+P^{ee}_{ij}(t,t')+P_{ij}^{hh}(t,t')$
and e.g.
\begin{eqnarray}
P_{ij}^{eh}(t,t')\propto \langle b_i^{e\dagger}(t)
b_j^{h\dagger}(t')b_j^{h}(t')b_i^{e}(t)\rangle.
\label{JDP}
\end{eqnarray}
The quasiparticle operators are defined as $b^e(t)=\int_0^{\infty}
dE~\mbox{exp}(-iEt/\hbar)b(E)$ and $b^h(t)=\int^0_{-\infty}
dE~\mbox{exp}(iEt/\hbar)b^{\dagger}(E)$. Evaluating the joint
detection probability, we find to leading order in $\delta V$ that
$P_{ij}\propto S_{ij}^{dc}$, as anticipated from the discussion below
Eq. (\ref{currcorr}). One can thus formulate a Bell inequality in
terms of the period-averaged, zero-frequency
noise.\cite{sam1,sam2,leb1,BI} Choosing an optimal\cite{sam1} set of
scattering angles $\{ \theta_A,\theta_B\}$ we arrive at the Bell
inequality $2\sqrt{1+\gamma^2}\leq 2$, maximally violated for
$\phi_D-\phi_C=\varphi=0~\mbox{mod}~2\pi$. Dephasing as well as
nonequal scattering potentials at $C$ and $D$ can be treated in the
same way as in Ref. [\onlinecite{sam1}].

So far we considered the limit of weak potential oscillations, where
only one quanta $\hbar \omega$ is absorbed or emitted by the
scattering electrons in regions $C$ and $D$. Relaxing this assumption,
for arbitrary strong potential modulations, it is no longer possible
to write the state emitted by contacts $C$ and $D$ as an excitation of
a single electron-hole pair out of the ground state. Instead, the
state can be written as a linear superposition of excitations of
multiple electron-pairs, describing a complicated multiparticle
entanglement. Moreover, while the amplitude of the weak potential
state oscillates with the single frequency $\omega$, the strong
potential state can have a complicated time-dependence with a sum of
amplitudes with oscillation frequencies $n\hbar \omega$.

It is nevertheless possible (here using the Floquet scattering theory
of Ref. [\onlinecite{Mosknoise}]) to calculate the current and
noise. Focusing on the dc-part, the current is zero for the same
reasons as in the weak potential case. Considering for simplicity
identical scattering potentials at $C$ and $D$, ($\phi_C=\phi_D$ as
well), the noise is given by the same expression as in
Eq. (\ref{currcorr2}), with
\begin{eqnarray}
\frac{hS_0}{2e^2}=\int dE \sum_{p} \left|\sum_{E_n<0}\left[t_{0-n}^*r_{p-n}+r_{0-n}'^{*}t'_{p-n}\right]\right|^2 
\label{currcorr3} 
\end{eqnarray}
where $t_{m-n}\equiv t(E_m,E_n)$ etc, with
$E_n=E+n\hbar\omega$. Interestingly, the strong potential modulation
only modifies the prefactor of the cross-correlations, not the
dependence on the angles $\theta_A,\theta_B$. This suggests that
two-particle entanglement might be postselected\cite{sam2,been2} by
the noise measurement itself. However, a calculation of the joint
detection probability $P_{ij}$ shows that it can in the general case,
i.e. without considering particular scattering potentials, not be
expressed in terms of the noise cross correlator in
Eq. (\ref{currcorr2}). This is a consequence of the complicated
time-dependence of the emitted state. Moreover, averaging the
time-dependent probability $P_{ij}(t,t')$ over a time much
longer than the period $\tau$ gives a $P_{ij}$ dominated by a
quasiparticle current product term, which makes a violation of the
Bell Inequality impossible.\cite{BI,leb1} It is not discussed here
whether there are specific strong pumping potentials for which
$P_{ij}\propto S^{dc}_{ij}$, allowing a formulation of a Bell
Inequality in terms of zero frequency noise.

In conclusion, we have proposed a scheme for generating quasi-particle
entanglement by a time-dependent scattering potential. The
entanglement is detected by violation of a Bell inequality, formulated
in terms of zero-frequency noise. This adiabatic generation of
orbitally entangled electron-hole pairs is considered both for weak
and strong potential modulations.

We acknowledge discussions with E. Sukhorukov, M. Polianski and
M. Moskalets. This work was supported by the Swiss NSF and the network
MaNEP.


\begin{thebibliography}{02}
\bibitem{Moskcom}
This electron-hole pair picture was discussed by Moskalets and
B\"uttiker in Refs. [\onlinecite{Mosk,Mosknoise}].
\end{thebibliography}
\end{document}